\documentclass[aps,prl,showpacs,twocolumn,floats,epsfig,pdflatex]{revtex4}
\usepackage{amssymb}
\usepackage{amsbsy}
\usepackage{amsmath}
\usepackage{epsfig}
\usepackage{amsfonts}
\usepackage{graphicx}
\usepackage{amssymb}
\usepackage{upgreek}
\usepackage{dsfont}
\usepackage{bm}
\usepackage{color}
\usepackage{hyperref}

\begin{document}

\title {Detecting band inversions by measuring the environment: 
fingerprints of electronic band topology in bulk phonon linewidths}
\author{Kush Saha}

\author{Katherine L\'egar\'e}

\author{Ion Garate}

\affiliation{D\'epartement de Physique and Regroupement Qu\'eb\'ecois sur les Mat\'eriaux de Pointe, Universit\'e de Sherbrooke, Sherbrooke, Qu\'ebec, Canada J1K 2R1}
\date{\today}
\begin{abstract}
The interplay between topological phases of matter and dissipative baths constitutes an emergent research topic with links to condensed matter, photonic crystals, cold atomic gases and quantum information.
While recent studies have suggested that dissipative baths can induce topological phases in intrinsically trivial quantum materials, the back action of topological invariants on dissipative baths has been overlooked.
By exploring this back action for a centrosymmetric Dirac insulator coupled to phonons, we show that the linewidths of bulk optical phonons can reveal electronic band inversions.
This result is the first known example where topological phases of an open quantum system may be detected by measuring the {\em bulk} properties of the surrounding environment.

\end{abstract}
\maketitle

{\em Introduction.--}
The discovery of topological phases in three dimensional crystals has culminated in a new classification scheme for solids that is based on quantum mechanics and topology~\cite{ti}.
These phases are described by integers known as topological invariants,
which manifest themselves through robust gapless states localized at sample boundaries.
The characterization of topological invariants in insulators often idealizes electrons as being isolated from their environment.
Yet, in real materials, electrons are coupled to various non-electronic baths
and the usual idealization fails when the strength of the coupling exceeds the energy gap of the insulator.

Recent work~\cite{diehl,garate}
has suggested that baths can alter topological invariants and even induce topological phases.
However, the inverse of this effect, concerning the back action of topological invariants on baths, remains completely unexplored. 
Does a change in the electronic topological invariant modify the surrounding bath?
Is it possible to infer the topological invariants of an electronic system by measuring its non-electronic environment? 
The present work intends to answer these questions affirmatively and thus establish an unanticipated interplay between band topology and dissipative baths.

To that end, we adopt a minimal model which consists of massive 3D Dirac fermions coupled to a bath of phonons.
In this model, we find that it is possible to learn whether the electronic band topology is trivial or nontrivial by analyzing the phonon linewidths in the thermodynamic limit (i.e. disregarding boundary effects).
Our results challenge a commonly held viewpoint, according to which the bulk properties of a doped topological insulator and a doped trivial insulator should be qualitatively similar.

\begin{figure}
\includegraphics[width=0.9\linewidth]{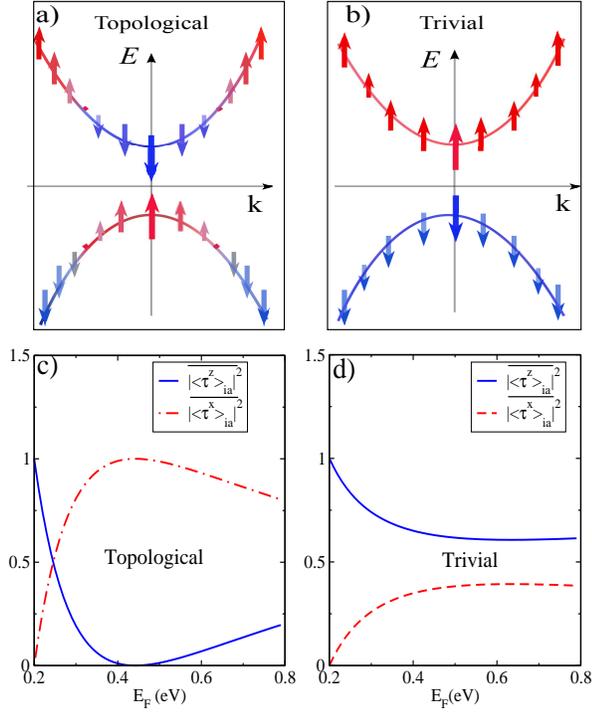} 
\caption{(Color online)  (a) and (b) Expectation value of the electronic parity operator, $\langle \tau^z\rangle$ (represented by arrows), as a function of momentum for the electronic model of Eq.~(\ref{eq:hm0}). 
(c) and (d) Fermi surface averages of $|\langle \tau^z\rangle|^2$ and $|\langle\tau^x\rangle|^2$
(cf. Eq.~(\ref{eq:tauxz})), as a function of the Fermi energy, for $m=-0.2\,{\rm eV}$ (c) and $m=0.2\,{\rm eV}$ (d).
The rest of the band parameters are the same as those in Ref.~\cite{garate}.}
\label{fig:fig1} 
\end{figure}

 {\em Model.--}
The minimal Hamiltonian describing the low-energy {\em bulk} bands of a time- and inversion-symmetric 3D Dirac insulator near the Brillouin zone center is~\cite{liu}
\begin{align}
\label{eq:hm0}
{\cal H}({\bf k})=\gamma k^2 +\alpha {\bf k}\cdot{\boldsymbol\sigma}\tau^x + M_{\bf k} \tau^z, 
\end{align}
where $\sigma^i$ and $\tau^i$ are Pauli matrices in spin and orbital space (respectively), ${\bf k}=(k_x,k_y,k_z)$ is the crystal momentum,  $\gamma$ models the particle-hole asymmetry of the band structure,  $\alpha$ is the Dirac velocity, $M_{\bf k}=m+\beta k^2$ is the Dirac mass, $2 |m|$ is the energy gap of the insulator at $k=0$,  and $\beta$ is an additional band parameter.
Importantly, $\tau^z$ is the electronic parity operator and $[\tau^z,{\cal H}({\bf 0})]=0$.
For narrow-gap insulators described by Eq.~(\ref{eq:hm0}), the sign of $m \beta$ determines the so-called strong topological invariant: if $\beta>0$, then $m>0$ ($m<0$) results in a trivial (topological) insulator.
If $m \beta<0$, the electronic bands at $k=0$ are said to be {\em inverted}.
In addition, $M_{\bf k}$ acts as a momentum-dependent effective magnetic field that polarizes the orbital pseudospin ${\boldsymbol\tau}$ along the $z$ direction. 
Because $M_{\bf k}$ changes sign as a function of $k$ in the topological phase but not in the trivial phase, the $k$-dependence of the expectation value of $\tau^z$ reflects the key difference between the {\em bulk} electronic structures of trivial and topological insulators (cf. Figs.~\ref{fig:fig1}a and \ref{fig:fig1}b).

The eigenstates of Eq.~(\ref{eq:hm0}) are $|u_{{\bf k} n}\rangle$,
where $n\in\{1,...,4\}$ labels the two highest valence bands and the two lowest conduction bands near $k=0$.
The energy eigenvalues, $E_{{\bf k} n}$, are doubly degenerate owing to the combined time-reversal and inversion symmetries.  
For analytical simplicity, we have chosen a continuum model with spherical symmetry; this captures the essential ideas and is smoothly deformable into more realistic lattice models that we use below for numerical calculations.

{\em Phonon self-energy.--}
Electron-phonon interactions shift phonon frequencies and contribute to phonon linewidth.
These two effects can be calculated from the phonon self-energy~\cite{ando},
\begin{equation}
\label{eq:pi}
\Pi_\lambda({\bf q},\omega_{{\bf q}\lambda})=\frac{1}{\cal V}\sum_{{\bf k} nn'} \frac{|g^\lambda_{nn'}({\bf k,q})|^2\left(f_{{\bf k}n}-f_{{\bf k-q}n'}\right)}
{E_{{\bf k}n}-E_{{\bf k-q}n'}-\omega_{{\bf q}\lambda}-i 0^+}.
\end{equation}
Here, ${\cal V}$ is the sample volume, $\lambda$ labels different phonon modes, ${\bf q}$ is the phonon momentum, $\omega_{{\bf q}\lambda}$ is the bare phonon frequency and  $f_{{\bf k}n}$ is the fermion occupation number for the state $|u_{{\bf k}n}\rangle$ with a Fermi energy $\epsilon_F$.
Also,
\begin{equation}
\label{eq:mat}
g^\lambda_{nn'}({\bf k,q})=\langle u_{{\bf k}n} | {\hat g}^\lambda ({\bf q})|u_{{\bf k-q} n'}\rangle,
\end{equation} 
where ${\hat g}^\lambda({\bf q})={\hat g}^\lambda(-{\bf q})^\dagger$ is the electron-phonon vertex operator in the low-energy electronic subspace~\cite{sm,local}.  

In a centrosymmetric crystal, lattice vibrations are either even or odd under spatial inversion.
Each of these modes couples to electrons and can, as we shall see,  inherit signatures of the underlying band topology.
For the model of Eq.~(\ref{eq:hm0}) and for $q\simeq 0$ optical phonons, inversion and time-reversal symmetries dictate~\cite{sm}
\begin{align}
\label{eq:ep}
{\hat g }^{\rm even}({\bf q}) & \simeq g_0 ({\bf {\hat q}}) +g_z ({\bf {\hat q}})\tau^{z}\nonumber\\
{\hat g}^{\rm odd}({\bf q})&\simeq g_x ({\bf {\hat q}})  \tau^{x}+{\bf g}'({\bf {\hat q}}) \cdot{\boldsymbol\sigma}\tau^{y},
\end{align}
where ``even'' (``odd'') denotes the coupling of electrons to parity-even (parity-odd) phonon modes, with $[{\hat g}^{\rm even},\tau^z]=0$ and $\{{\hat g}^{\rm odd},\tau^z\}=0$.
Inversion symmetry guarantees that ${\hat g}^{\rm even}$ and ${\hat g}^{\rm odd}$ will not be mixed in a single phonon mode.
Also, $\hat{\bf q}={\bf q}/q$, and 
the coefficients $g_i$ ($i=0,x,z$) and $g'_i$ ($i=x,y,z$) can be obtained from the atomic displacements in the particular phonon mode~\cite{sm}.
Physically, $g_0$ and $g_z$ lead to phonon-induced modulations of the chemical potential and the Dirac mass, respectively. 
Next, we identify ways in which ${\hat g}^\lambda$ can transfer the information about electronic band topology to the phonon sector.

{\em Intraband phonon damping.--}
The main electronic mechanism contributing to phonon linewidths is the scattering of phonons off electron-hole pairs.
The rate of this process is $\gamma^\lambda({\bf q})\equiv-{\rm Im}\,\Pi_\lambda({\bf q},\omega_{{\bf q} \lambda})$.
In this work, we focus on long-wavelength optical phonons and on low temperatures.

We begin by considering the commonly realized case in which the phonon frequency is smaller than the bandgap of the insulator. 
In this case, the ``insulator'' must be doped in order for carriers to absorb phonons and induce a linewidth $\gamma^\lambda_{\rm ia}$. 
The subscript ``${\rm ia}$'' is shorthand for ``intraband'' and makes it explicit that phonons decay into particle-hole pairs in the vicinity of the Fermi surface.
Assuming that the distance from the Fermi level to the bulk band edge is large compared to the phonon frequency, we have~\cite{sm}
\begin{equation}
\gamma^\lambda_{\rm ia} (q\simeq 0) \simeq \pi \omega_{{\bf 0}\lambda} D(\epsilon_F) \overline{|g^\lambda_{\rm ia}({\bf k}_F, \hat{\bf q})|^2 \delta({\bf v}_F\cdot{\bf q}-\omega_{{\bf 0}\lambda})},
\label{eq:img}
\end{equation}
where $D(\epsilon_F)$ is the electronic density of states per band at the Fermi level, ${\bf k}_F$ is the Fermi wave vector,  ${\bf v}_F$ is the Fermi velocity, $\delta (x)$ is the Dirac delta, and
$|g^\lambda_{\rm ia}({\bf k},\hat{\bf q})|^2$ denotes the sum of $|g^\lambda_{n n'}|^2$ over the two degenerate bands at momentum ${\bf k}$ and energy $E_{{\bf k}}$ (hence the label ``intraband'').
In addition, 
$\overline{O}=\sum_{{\bf k}} O\delta(E_{{\bf k}}-\epsilon_F)/({\cal V} D(\epsilon_F))$.

Equation~(\ref{eq:img}) contains information about the electronic band topology. 
The simplest way to see this is to imagine a parity-even phonon mode and a parity-odd phonon mode that couple to electrons purely through ${\hat g}^{\rm z}\equiv g_z \tau^z$ and ${\hat g}^{\rm x}\equiv g_x \tau^x$, respectively. 
More general couplings with $g_0\neq 0$ and $g'_i\neq 0$ will be discussed below.
From Eqs.~(\ref{eq:hm0}), (\ref{eq:mat}) and (\ref{eq:img}), the linewidths of these two phonon modes are~\cite{sm} 
\begin{align}
\label{eq:imgi}
\gamma^j_{\rm ia}(q\simeq 0) &\simeq |g_j(\hat{\bf q})|^2 D(\epsilon_F) \overline{|\langle\tau^j\rangle_{\rm ia}|^2} \frac{\pi \eta}{2} \Theta(1-\eta),
\end{align}
where $j\in\{x,z\}$, $\Theta(x)$ is the Heaviside function, $\eta\equiv \omega_{{\bf 0}j}/ (q v_F)$ and
\begin{equation}
\label{eq:tauxz}
\overline{|\langle\tau^z\rangle_{\rm ia}|^2} = 1-\overline{|\langle\tau^x\rangle_{\rm ia}|^2}= M_{k_F}^2/(\alpha^2 k_F^2 + M_{k_F}^2). 
\end{equation} 
Note that $\overline{|\langle\tau^j\rangle_{\rm ia}|^2}\in[0,1]$ (cf. Fig.~\ref{fig:fig1}).
In particular, when $M_{k_F}=0$, $\overline{|\langle\tau^z\rangle_{\rm ia}|^2}=0$ and $\overline{|\langle\tau^x\rangle_{\rm ia}|^2}=1$.
Combining Eqs.~(\ref{eq:imgi}) and (\ref{eq:tauxz}) with Fig.~\ref{fig:fig1}, it follows that $\gamma^j_{\rm ia}$ reflects the orbital texture, and therefore the topology, of the bulk bands.
In order to clarify this point, we eliminate the non-topological features coming from $D(\epsilon_F)$ by considering the ratio $\gamma^x_{\rm ia}/\gamma^z_{\rm ia}\simeq (g_x^2/g_z^2) \overline{|\langle\tau^x\rangle_{\rm ia}|^2}/\overline{|\langle\tau^z\rangle_{\rm ia}|^2}$.

\begin{figure}
\includegraphics*[width=0.95\linewidth]{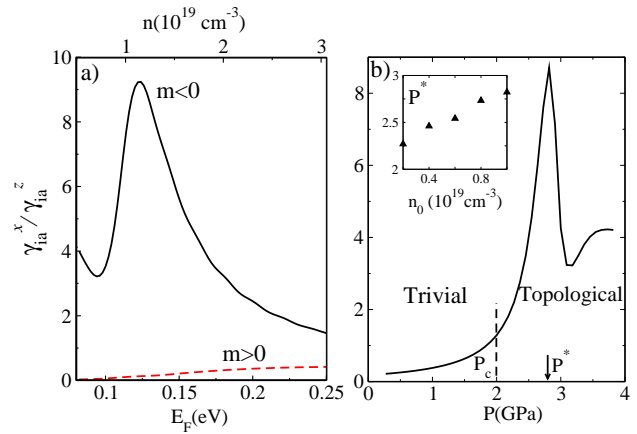}
\caption{
(Color online) (a) $\gamma^x_{\rm ia}/\gamma^z_{\rm ia}$ as a function of the Fermi energy and the bulk carrier density, for $m=\pm 0.1\,{\rm eV}$.   
A prominent maximum emerges in the topological phase only, due to the momentum-space orbital texture of the electronic eigenstates.
(b) $\gamma^x_{\rm ia}/\gamma^z_{\rm ia}$ as a function of pressure $P$.
We use $m=\alpha (P-P_c)$, where $P_c$ is the critical pressure for a band inversion and $\alpha$ is a coefficient that can be obtained e.g. from experiment~\cite{bera}.
The bulk carrier density is  $n\simeq n_0(1+P/B)$, where $n_0$ is the density at $P=0$ and $B$ is the bulk modulus.
The maximum of $\gamma^x_{\rm ia}/\gamma^z_{\rm ia}$ appears at $P=P^*$.
{\it Inset}: The dependence of $P^*$ on $n_0$. 
As $n_0$ decreases, $P^*$ approaches $P_c$, making it more difficult to identify trivial and topological phases solely from phonon measurements. 
Throughout this figure, we have used a tetragonal lattice regularization of Eq.~(\ref{eq:hm0}).
Because $\alpha,\beta,\gamma$ are not tabulated for Sb$_2$Se$_3$, we have replaced them with those of Sb$_2$Te$_3$~\cite{liu}.
For the bulk modulus, we have used $B=30\,{\rm GPa}$~\cite{caution}.
}
\label{fig:fig2}
\end{figure}

For fixed bandgap, Eq.~(\ref{eq:imgi}) predicts a strong maximum for $\gamma^x_{\rm ia}/\gamma^z_{\rm ia}$ as a function of $\epsilon_F$ in the topological phase (but not in the trivial phase) because $M_{k_F}$ crosses zero as a function of $\epsilon_F$ in the topological phase (but not in the trivial phase).
This difference in behavior between the trivial and topological phases is significant for a sizeable $|m|$, but becomes gradually weaker as the energy gap decreases, ultimately disappearing when $m\to 0$. 
In other words, $\gamma^x_{\rm ia}/\gamma^z_{\rm ia}$ contains 
no signatures of band topology near the topological quantum critical point.
Figure~\ref{fig:fig2}a confirms our analytical statements in a lattice model, for which Eq.~(\ref{eq:img}) is solved numerically.

Alternatively, in a sample with fixed carrier density, $\gamma^x_{\rm ia}/\gamma^z_{\rm ia}$ shows a pronounced maximum as a function of $m$ in the topological phase only.
The maximum takes place at $m^*\simeq -\beta k_F^2$, where $M_{k_F}$ undergoes a sign change.
Motivated by recent claims of pressure-induced band inversions in Sb$_2$Se$_3$ and Pb$_{1-x}$Sn$_x$Se~\cite{bera, xi},
in Fig.~\ref{fig:fig2}b we plot $\gamma^x_{\rm ia}/\gamma^z_{\rm ia}$ as a function of pressure, using a lattice model.
This corroborates the emergence of a ``topology-induced'' maximum in $\gamma^x_{\rm ia}/\gamma^z_{\rm ia}$.

In the preceding discussion of $\gamma^x_{\rm ia}$, we have assumed a parity-odd phonon mode that couples to electrons purely through $\tau^x$ ($g'_i=0$ in Eq.~(\ref{eq:ep})).
In general, such a phonon can also couple to electrons through the term ${\bf g}'\cdot{\boldsymbol\sigma}\tau^y$. 
However, we have verified that this coupling produces qualitatively similar features as $\tau^x$.

Similarly, when discussing $\gamma^z_{\rm ia}$, we have imagined a parity-even phonon mode that couples to electrons purely through $\tau^z$ ($g_0=0$ in Eq.~(\ref{eq:ep})).
Nonetheless, symmetry allows a mixture of $\tau^z$ and the identity matrix ${\bf 1}$~\cite{ph}.
The latter produces intraband matrix elements that are insensitive to the orbital texture of the insulator, since $\langle u_{{\bf k} n}| {\bf 1}|u_{{\bf k} n}\rangle=1$.
Consequently, the effect of $g_0\neq 0$ is to dilute away the topological features of $\gamma^z_{\rm ia}$.
Although this constitutes a problem towards the realization of Fig.~\ref{fig:fig2} in real materials, we find that the maximum in $\gamma^x_{\rm ia}/\gamma^z_{\rm ia}$ remains pinned to the topological side if $|g_z|>|g_0|$.

{\em Interband phonon damping.--}
Thus far, we have considered the linewidths of phonons with $\omega_{{\bf 0} \lambda}<2 |m|$. 
Herein, we investigate the case $\omega_{{\bf 0} \lambda} > 2 |m|$, relevant to Dirac insulators with particularly small bandgaps and/or high-frequency phonon modes.  
In this case, a phonon is absorbed by an electron in the bulk valence band, which gets promoted to the bulk conduction band.
The associated phonon linewidth is $\gamma^\lambda_{\rm ie}$, where the subscript ``ie'' is shorthand for ``interband''.
Assuming for the moment that $\epsilon_F$ is inside the bulk gap, Eq.~(\ref{eq:pi}) yields~\cite{sm}
\begin{equation}
\gamma^\lambda_{\rm ie}(q\simeq 0) \simeq \pi D_{\rm joint}(\omega_{{\bf 0} \lambda}) \overline{|g^\lambda_{\rm ie}({\bf k}, \hat {\bf q})|^2},
\label{eq:img3}
\end{equation}
where $D_{\rm joint}(\omega)=\sum_{{\bf k}} \delta(E_{{\bf k} c}-E_{{\bf k} v} - \omega)/{\cal V}$ is the joint density of states, $E_{{\bf k} c}$ and $E_{{\bf k} v}$ are the bulk conduction (c)  and valence (v) band energies. 
In addition, $|g^\lambda_{\rm ie}|^2=\sum_{n\in{\rm c}, n'\in{\rm v}} |g^\lambda_{n n'}|^2$ and
$\overline{|g^\lambda_{\rm ie}|^2}=\sum_{\bf k} |g^\lambda_{\rm ie}|^2\delta(E_{{\bf k}c}-E_{{\bf k}v}-\omega_{{\bf 0} \lambda})/({\cal V} D_{\rm joint})$.

\begin{figure}
\includegraphics*[width=0.9\linewidth]{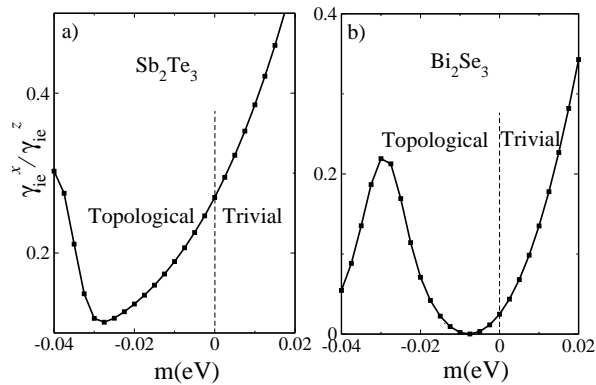}
\caption{
(Color online) 
$\gamma^x_{\rm ie}/\gamma^z_{\rm ie}$ as a function of the Dirac mass $m$, where the rest of the band parameters correspond to Sb$_2$Te$_3$ (a) or Bi$_2$Se$_3$ (b).
The minimum  of $\gamma^x_{\rm ie}/\gamma^z_{\rm ie}$ occuring in the topological side is a direct manifestation of the orbital texture in Fig.~\ref{fig:fig1}.
Throughout this figure, we have used a tetragonal lattice regularization of Eq.~(\ref{eq:hm0}) with the band parameters taken from Ref.~[\onlinecite{liu}]. 
The Fermi level is assumed to be inside the bulk gap.
}
\label{fig:fig3}
\end{figure}

From Eqs.~(\ref{eq:hm0}) and (\ref{eq:mat}), we obtain $|g^z_{\rm ie}({\bf k},0)|^2=|g^x_{\rm ia}({\bf k},0)|^2$ and $|g^x_{\rm ie}({\bf k},0)|^2=|g^z_{\rm ia}({\bf k},0)|^2$.
Therefore, $\gamma^\lambda_{\rm ie}$ is as sensitive as $\gamma^\lambda_{\rm ia}$ to the band topology of the Dirac insulator (with $x$ and $z$ interchanged).
More so, an important advantage of $\gamma^\lambda_{\rm ie}$ over $\gamma^\lambda_{\rm ia}$ is that we may effectively take
$\hat{g}^{\rm even} = \hat{g}^z$ {\em regardless} of the value of  $g_0$ in Eq.~(\ref{eq:ep}), because $\langle u_{{\bf k} n}| {\bf 1} | u_{{\bf k} n'}\rangle=0$ for interband transitions.
Accordingly, the topological signatures in $\gamma_{\rm ie}^\lambda$ are more robust than those in $\gamma^\lambda_{\rm ia}$.

In a sample with fixed carrier density, $\gamma^x_{\rm ie}/\gamma^z_{\rm ie}$ contains a {\em minimum} as a function of $m$ at $m^*\simeq -\omega_0^2 \beta/(4 \alpha^2)$, i.e. only in the topological phase~\cite{caveat}.
This result has the same origin as the maximum of $\gamma^x_{\rm ia}/\gamma^z_{\rm ia}$ discussed above, and it holds for doped samples as well so long as $\alpha k_F/\omega_0\ll 1$.
Figure~\ref{fig:fig3} confirms this for a lattice model.

{\em Discussion.--}
In sum, there are three reasons why the linewidths of bulk, long-wavelength optical phonons can inherit distinct signatures of the electronic band topology.
First, phonon linewidths are proportional to the square of energy-resolved electron-phonon matrix elements.
Second, in a centrosymmetric crystal, the coupling of a optical $q\simeq 0$ phonon to electrons either commutes or anticommutes with the electronic parity operator.
Third, the momentum-space texture of electronic parity eigenvalues 
reflects band inversions.


To be precise, the predicted features in the phonon linewidths probe {\em band inversions} near the Fermi level, rather than the strong topological invariant {\em per se}.
However, in materials whose low-energy bands are described by Eq.~(\ref{eq:hm0}), the energy gap minimum occurs at a single time-reversal-invariant momentum (TRIM) and hence a band inversion taking place therein 
is equivalent to a change in the strong topological invariant.
By extension, phonon linewidths can probe the strong topological invariant in centrosymmetric crystals where the direct bandgap minimum is known to occur at an {\em odd} number of symmetry-equivalent TRIM.
In contrast, phonon linewidths do not faithfully reflect the strong topological invariant in materials where the direct bandgap minimum can occur at an even number of symmetry-equivalent TRIM, nor in materials lacking an inversion center.

Phonon frequencies, which we have barely  mentioned thus far, are  much less sensitive than phonon linewidths to the electronic band topology. 
This is because the real part of Eq.~(\ref{eq:pi}) contains a sum over electron-phonon matrix elements at multiple energies, with weights that depend on non-topological details of the energy bands.
This notwithstanding, a recent experiment~\cite{bera} in Sb$_2$Se$_3$ has attributed a kink in the pressure-dependence of the phonon frequency
to a band inversion. 
Our calculations~\cite{sm} 
do not support such interpretation. 

The main tools to measure $q\simeq 0$ phonon linewidths are Raman spectroscopy (for parity-even phonons), infrared spectroscopy (for parity-odd phonons) and inelastic neutron scattering~\cite{cardona}.
In a clean material with $\omega_{{\bf 0} \lambda}\tau\gg 1$ (where $\tau$ is the disorder scattering time), $\gamma^\lambda_{\rm ia}$ vanishes unless $q>\omega_{{\bf 0}\lambda}/v_F$~\cite{phonq}.
Since $\omega_{{\bf 0}\lambda}/v_F$ typically exceeds the photon wave vector used in optical spectroscopies, $\gamma^\lambda_{\rm ia}$ should be measured with neutrons. 
In contrast, $\gamma^\lambda_{\rm ie}$ remains nonzero at $q=0$ and is thus amenable to optics.
For 
Bi$_2$Se$_3$, we estimate $\gamma^\lambda_{\text{ie, ia}}\lesssim 1\, {\rm cm}^{-1}$, which nears the experimental resolution~\cite{shapiro}.

Aside from electron-phonon interactions, anharmonic lattice effects contribute to the phonon linewidth.
To leading order, phonon-phonon interactions contain no information about the electronic band topology and are independent from the carrier density. 
Therefore, the anharmonic part can be subtracted by measuring the linewidths with respect to a baseline carrier density. 

In view of our results, it is natural to ask whether any other physical observable involving Fermi's golden rule, such as conductivity, might be sensitive to electronic band topology on the same footing as the phonon linewidths.
The answer is generally negative.
For example, the optical conductivity cannot clearly differentiate between trivial and nontrivial orbital textures in the bulk because the velocity operator mixes  ${\bf 1}$, $\tau^x$, and $\tau^z$~\cite{gnez}. 

To conclude, we have proven that it is in principle possible to infer the strong topological invariant of multiple Dirac insulators from the linewidths of bulk, long-wavelength optical phonons.
It may be of interest to investigate formal links between the phonon linewidth and the SU(2) Berry phase identified in Ref.~\cite{hosur}.
Likewise, it will be desirable to complement our theory with {\em ab-initio} electronic structure calculations, and to search for similar insights in other contexts (cold atoms, photonic crystals, quantum memories) where the interplay between topology and dissipation may be crucial.

{\em Acknowledgements.--}
We are grateful to K. Pal and U. Waghmare for sharing useful information about Ref.~[\onlinecite{bera}].
This work has been funded by Qu\'ebec's RQMP and Canada's NSERC.
The numerical calculations were performed on computers provided by Calcul Qu\'ebec and Compute Canada.

\begin{widetext}
\section{Supplementary material}
%
%


\subsection{Electron-phonon coupling}
The objective of this section is to derive Eq.~(4) of the main text. 
For the first part of the derivation, we follow Ref.~[\onlinecite{allen}].
We start with the periodic lattice potential given by
\begin{align}
U({\bf r})=\sum_{{\bf l},\beta} U_{\beta}({\bf r}-{\bf R}({\bf l},\beta)), 
\label{eq:pot}
\end{align}
where ${\bf R}(l,\beta)={\bf l}+{\boldsymbol \tau}_{\beta}$ is the equilibrium position of an atom $\beta$ located in the $l-$th unit cell, 
${\bf l}$ being the position of the unit cell and ${\boldsymbol \tau}_{\beta}$ being the position of the atom with respect to a reference point in that unit cell.
The departures from the atoms from their equilibrium positions produce a change in the lattice potential, 
\begin{align}
\delta U({\bf r })=&\sum_{{\bf l},\beta} U_{\beta}({\bf r}-{\bf R}({\bf l},\beta)-{\bf Q}({\bf l},\beta) )-\sum_{{\bf l},\beta} U_{\beta}({\bf r}-{\bf R}({\bf l},\beta))\nonumber\\
&\simeq \sum_{{\bf l},\beta}{\bf Q}({\bf l},\beta)\cdot{\partial U_{\beta}({\bf r}-{\bf R}({\bf l},\beta)) \over \partial {\bf R}({\bf l},\beta)},
\label{eq:pot1}
\end{align} 
where ${\bf Q}({\bf l},\beta) $ is the displacement operator for atom $\beta$ in unit cell ${\bf l}$ away from its equilibrium position,  given by  
\begin{equation}
{\bf Q}({\bf l},\beta) =\sum_{{\bf q},\lambda} \sqrt{\hbar\over 2M_{\beta} N\omega_{{\bf q}\lambda}} e^{i {\bf q}\cdot({\bf l}+\boldsymbol{\tau}_{\beta})}{\bf p}_\lambda({\bf q},\beta) (a_{-{\bf q}\lambda}+a_{{\bf q}\lambda}^{\dagger}).
\label{eq:dis}
\end{equation}
Here, ${\bf q}$ is the phonon momentum, $a_{{\bf q}\lambda}$ is an operator that annihilates a phonon mode $\lambda$ with momentum ${\bf q}$, $M_\beta$ is the mass of atom $\beta$,  ${\bf p}_\lambda({\bf q},\beta)$ is the polarization vector corresponding to atom $\beta$ in mode $\lambda$ and momentum ${\bf q}$, $\omega_{{\bf q}\lambda}$ is the phonon frequency and $N$ is the number of unit cells in the crystal. 
 
Then, the local electron-phonon coupling can be written as
\begin{align}
{\cal H}^{\rm loc}_{\rm ep}=\int d{\bf r}\,  \rho({\bf r})\, \delta U({\bf r}),
\label{eq:hm1}
\end{align}
where $\rho ({\bf r})=\psi^{\dagger}({\bf r})\psi({\bf r})$  is the electron density operator, 
\begin{equation}
\psi({\bf r})=\frac{1}{\sqrt{\cal V}}\sum_{n\bf k} e^{-i {\bf k. r}} u_{{\bf k}n}({\bf r})c_{{\bf k}n},
\end{equation}
${\cal V}$ is the volume of the sample, $u_{{\bf k} n}({\bf r})$ is the periodic part of the Bloch wave function corresponding to band $n$ and momentum ${\bf k}$ (within the first Brillouin zone), and $c_{{\bf k}n}$ is an operator that annihilates an electron in such a state.
It follows that
 \begin{align}
 {\cal H}^{\rm loc}_{\rm ep}=&\frac{1}{\cal V}\sum_{\beta {\bf l}}\sum_{{\bf k}{\bf k}'}\sum_{n n'}\int d{\bf r}\, e^{i{\bf (k-k').r}} u_{{\bf k}n}^{\ast}({\bf r}) u_{{\bf k'}n'}({\bf r})\, {\bf Q}({\bf l},\beta)\cdot{\partial U_{\beta}({\bf r}-{\bf R}({\bf l},\beta) )\over \partial {\bf R}({\bf l},\beta)} c_{{\bf k}n}^{\dagger} c_{{\bf k'}n'}\nonumber\\
 =&\frac{1}{\cal V}\sum_{\beta {\bf l}}\sum_{{\bf k}{\bf k}'}\sum_{n n'}\int d{\bf r}\, e^{i{\bf (k-k').(r+l)}} u_{{\bf k}n}^{\ast}({\bf r+l}) u_{{\bf k'}n'}({\bf r+l})\, {\bf Q}({\bf l},\beta)\cdot{\partial U_{\beta}({\bf r+l}-{\bf R}({\bf l},\beta) )\over \partial {\bf R}({\bf l},\beta)}c_{{\bf k}n}^{\dagger}c_{{\bf k'}n'}\nonumber\\
 =&\frac{N}{\cal V}\sum_{\beta} \sum_{{\bf k}{\bf k}'}\sum_{n n'} \sum_{{\bf q}\lambda}\delta_{{\bf k'}, {\bf k+q}}\sqrt{\hbar\over 2 M_{\beta} N \omega_{{\bf q}\lambda}} \int d{\bf r}\, e^{i {\bf q}.(-{\bf r}+{\boldsymbol \tau}_{\beta})} u_{{\bf k}n}^{\ast}({\bf r}) u_{{\bf k'}n'}({\bf r})\, {\bf p}_\lambda({\bf q},\beta)\cdot{\partial U_{\beta}({\bf r}-{\bf R}({\bf 0},\beta)) \over \partial {\bf R}({\bf 0},\beta)} c_{{\bf k} n}^{\dagger}c_{{\bf k'} n'}(a_{-{\bf q}\lambda}+a_{{\bf q}\lambda}^{\dagger})\nonumber\\
=&\frac{1}{{\cal V}_{\rm cell}}\sum_{\beta} \sum_{\bf k}\sum_{n n'} \sum_{{\bf q}\lambda }\sqrt{\hbar\over 2 M_{\beta} N \omega_{{\bf q}\lambda}} \int d{\bf r} e^{i {\bf q}.(-{\bf r}+{\boldsymbol \tau}_{\beta})} u_{{\bf k}n}^{\ast}({\bf r})u_{{\bf k+q} n'}({\bf r})\, {\bf p}_\lambda({\bf q},\beta)\cdot{\partial U_{\beta}({\bf r}-{\bf R}({\bf 0},\beta)) \over \partial {\bf R}({\bf 0},\beta)}c_{{\bf k} n}^{\dagger}c_{{\bf k+q} n'}(a_{-{\bf q}\lambda}+a_{{\bf q}\lambda}^{\dagger})\nonumber\\
\equiv&\frac{1}{\sqrt{\cal V}}\sum_{{\bf k} n n'}\sum_{{\bf q}\lambda} g^\lambda_{nn'}({{\bf k,q}}) c_{{\bf k}n}^{\dagger}c_{{\bf k+q}n'}(a_{-{\bf q}\lambda}+a_{{\bf q}\lambda}^{\dagger}).
\label{eq:loc}
 \end{align}
where ${\cal V}_{\rm cell}={\cal V}/N$ is the unit cell volume and 
\begin{align}
g^\lambda_{n n'}({\bf k,q})\equiv&\frac{1}{\sqrt{{\cal V}_{\rm cell}}}\sum_{\beta}\sqrt{\hbar\over 2 M_{\beta} \omega_{{\bf q}\lambda}} \int d{\bf r}\, e^{i {\bf q}.(-{\bf r+}{\boldsymbol \tau}_{\beta})} u_{{\bf k}n}^{\ast}({\bf r}) u_{{\bf k+q}n'}({\bf r})\,{\bf p}_\lambda({\bf q},\beta)\cdot{\partial U_{\beta}({\bf r}-{\bf R}({\bf 0},\beta) )\over \partial {\bf R}({\bf 0},\beta)}\nonumber\\
&=\frac{1}{\sqrt{{\cal V}_{\rm cell}}}\sum_{\beta}\sqrt{\hbar\over 2 M_{\beta} \omega_{{\bf q}\lambda}} e^{i {\bf q}\cdot{\boldsymbol\tau}_\beta} {\bf p}_\lambda({\bf q},\beta)\cdot \langle u_{{\bf k} n}|e^{-i {\bf q .r}} {\partial U_{\beta}({\bf r}-{\bf R}({\bf 0},\beta) )\over \partial {\bf R}({\bf 0},\beta)} |u_{{\bf k+q}n'}\rangle 
\label{eq:matel}
\end{align}
is the matrix element that appears in Eqs.~(2) and (3) of the main text.
In Eq.~(\ref{eq:loc}), we have used $u({\bf r}+{\bf l})=u({\bf r})$ and $\sum_{\bf l} \exp(i {\bf k}\cdot{\bf l})= N \delta_{{\bf k},{\bf 0}}$.
It is implicit that, whenever ${\bf k}+{\bf q}$ exits the first Brillouin zone, a reciprocal lattice vector will be added to bring it back in.
The units of $g^\lambda_{n n'}$ are energy $\times$ $\sqrt{\text{volume}}$.
In spite of the integration over the entire crystal in Eq.~(\ref{eq:matel}), $g^\lambda_{n n'}$ is an intensive quantity (i.e. independent of the volume of the crystal in the thermodynamic limit).
This is a consequence of the fact that ${\partial U_{\beta}({\bf r}-{\bf R}({\bf 0},\beta) )/ \partial {\bf R}({\bf 0},\beta)}$ is localized in space, while the Bloch functions are extended.

Thus far everything has been general~\cite{allen}.
From now on, we will limit ourselves to the low-energy electronic bands spanned by the two highest valence bands and the two lowest conduction bands in the vicinity of the $\Gamma$ point.
Within this four-band subspace, it is useful to transform $g^\lambda_{n n'}({\bf k,q})$ into the basis spanned by $|u_{\sigma\tau}\rangle$, which are the low-energy electronic eigenstates at $\Gamma$ (i.e. at $k=0$).
Specifically,
\begin{equation}
|u_{{\bf k}n}\rangle=\sum_{\sigma\tau}|u_{\sigma\tau}\rangle\langle u_{\sigma\tau} | u_{{\bf k}n}\rangle,
\end{equation}
where $\sigma$ and $\tau$ are the two-level degrees of freedom describing spin and orbital, respectively, and $\langle u_{\sigma\tau}|u_{{\bf k} n}\rangle$ can be obtained by diagonalizing a ${\bf k}\cdot{\bf p}$ Hamiltonian (e.g. Eq.~(1) from the main text).
Then,
\begin{align}
g^\lambda_{n n'}({\bf k,q})
&=\sum_{\sigma\tau}\sum_{\sigma'\tau'}   \langle u_{{\bf k}n}| u_{\sigma\tau}\rangle \langle u_{\sigma'\tau'}|u_{{\bf k+q}n'}\rangle g^\lambda_{\sigma\tau;\sigma'\tau'}({\bf q}),
\label{eq:matel1}
\end{align}
where 
\begin{equation}
\label{eq:gst}
g^\lambda_{\sigma\tau;\sigma'\tau'}({\bf q})\equiv\frac{1}{\sqrt{{\cal V}_{\rm cell}}}\sum_{\beta}\sqrt{\hbar\over 2 M_{\beta} \omega_{{\bf q}\lambda}} e^{i{\bf q}\cdot{\boldsymbol\tau}_\beta}  \int d{\bf r} e^{-i {\bf q.r}}  u_{\sigma\tau}({\bf r})^* u_{\sigma'\tau'}({\bf r})\, {\bf p}_\lambda({\bf q},\beta)\cdot{\partial U_{\beta}({\bf r}-{\bf R}({\bf 0},\beta) )\over \partial {\bf R}({\bf 0},\beta)}.
\end{equation}
It is now possible to rewrite ${\cal H}^{\rm loc}_{\rm ep}$ in the basis spanned by $\sigma$ and $\tau$:
\begin{equation}
{\cal H}^{\rm loc}_{\rm ep}=\frac{1}{\sqrt{{\cal V}}}\sum_{{\bf k}}\sum_{{\bf q}\lambda}\sum_{\sigma\tau}\sum_{\sigma'\tau'} g^\lambda_{\sigma\tau; \sigma'\tau'}({\bf q})\, c_{{\bf k}\sigma\tau}^{\dagger} c_{{\bf k+q}\sigma'\tau'}(a_{-{\bf q}\lambda}+a_{{\bf q}\lambda}^{\dagger}).
\end{equation}
In general, the evaluation of $g^\lambda_{\sigma\tau; \sigma'\tau'}({\bf q})$ requires a detailed knowledge of the crystal structure of the material and of the electronic wave functions at $\Gamma$.
Here, we are interested in making some generic statements based on symmetries.
Let $\hat{g}^\lambda({\bf q})$ be a $4\times 4$ matrix such that
\begin{equation}
\langle u_{\sigma\tau}| \hat{g}^\lambda({\bf q}) | u_{\sigma'\tau'}\rangle \equiv g_{\sigma\tau; \sigma'\tau'}^\lambda ({\bf q}).
\end{equation}
When the matrix $\hat{g}^\lambda ({\bf q})$ acts on $|u_{\sigma\tau}\rangle$, the latter can be thought of as a 4-component spinor. 
The most general form for $\hat{g}^\lambda ({\bf q})$ is
\begin{equation}
\label{eq:ghat}
\hat{g}^\lambda ({\bf q})=\sum_{i j} g^\lambda_{ij}({\bf q}) \sigma^i \tau^j,
\end{equation}
where $i,j\in\{0,x,y,z\}$, the $0$-th Pauli matrix represents the identity, and $g^\lambda_{i j}({\bf q})$ are complex numbers.
There are various constraints on the form of the coefficients $g_{i j}^\lambda$.
First, since ${\cal H}_{\rm ep}^{\rm loc}$ is Hermitian, the matrix $\hat{g}$ must obey
\begin{equation}
\label{eq:gher}
\hat{g}^\lambda({\bf q}) = \hat{g}^\lambda(-{\bf q})^\dagger.
\end{equation}
Second, time-reversal symmetry dictates
\begin{equation}
\label{eq:gtim}
\Theta^{-1} \hat{g}^\lambda({\bf q}) \Theta = \hat{g}^\lambda(-{\bf q}), 
\end{equation}
where $\Theta=i \sigma^y K$ is the time-reversal operator and $K$ is the complex-conjugate operator.
By combining Eqs.~(\ref{eq:gher}) and (\ref{eq:gtim}), it follows immediately that
\begin{equation}
\label{eq:trev}
 \text{\em only matrices that commute with $\Theta$ are allowed in the expansion of $\hat{g}$}.
\end{equation}
This rules out many terms in Eq.~(\ref{eq:ghat}).  
Yet, as we show next, if the crystal has inversion symmetry, there are additional constraints in the limit of long-wavelength phonons ($q\simeq 0$), which simplify the form of $\hat{g}$ considerably.
When $q\simeq 0$, Eq.~(\ref{eq:gst}) may be simplified through 
\begin{equation}
\label{eq:exp1}
e^{i {\bf q}\cdot{\boldsymbol\tau}_\beta} \simeq 1
\end{equation}
and
\begin{equation}
\label{eq:exp2}
e^{-i {\bf q}\cdot{\bf r}}\simeq 1 - i {\bf q}\cdot{\bf r} + ...
\end{equation}
In Eq.~(\ref{eq:exp1}), the Taylor expansion has been truncated at the leading term because the size of a unit cell is negligible compared to the phonon wavelength when $q\simeq 0$.
The expansion in Eq.~(\ref{eq:exp2}) is at first glance less straightforward, as ${\bf r}$ can be arbitrarily large.
However, let us recognize the fact that $\partial U({\bf r}-{\bf R}({\bf 0},\beta))/\partial {\bf R}({\bf 0},\beta)$ is a function that is localized about the ${\bf l}={\bf 0}$ unit cell.
Then, the main contribution to the integral in Eq.~(\ref{eq:gst}) comes from ${\bf r}$ in the ``vicinity'' of the ${\bf l}={\bf 0}$ unit cell, where ``vicinity'' is quantified by the
range of $\partial U({\bf r}-{\bf R}({\bf 0},\beta))/\partial {\bf R}({\bf 0},\beta)$.
For $q\simeq 0$ phonons, this range is always short compared to the phonon wavelength, which then justifies the expansion in Eq.~(\ref{eq:exp2}).
Moreover, for $q\simeq 0$ optical phonons, it is sufficient to keep only the leading term in Eq.~(\ref{eq:exp2}).
A similar type of argument has been invoked in Ref.~[\onlinecite{frank}].

The main advantage of considering the long-wavelength limit in a centrosymmetric crystal is the emergence of a simple parity selection rule.
Let us begin by writing Eq.~(\ref{eq:gst}) as
\begin{align}
g^\lambda_{\sigma\tau;\sigma'\tau'}(\hat{\bf q})\simeq &\sum_{\beta}\frac{1}{\sqrt{{\cal V}_{\rm cell}}}\sqrt{\hbar\over 2 M_{\beta}  \omega_{{\bf q}\lambda}}  \int d{\bf r}\, u^{\ast} _{\sigma\tau}({\bf r})u_{\sigma'\tau'} ({\bf r}) {\bf p}_\lambda(\hat{\bf q},\beta)\cdot{\partial U_{\beta}({\bf r}-{\bf R}({\bf 0},\beta) )\over \partial {\bf R}({\bf 0},\beta)}\nonumber\\
=&\sum_{\beta}\frac{1}{\sqrt{{\cal V}_{\rm cell}}}\sqrt{\hbar\over 2 M_{\beta}  \omega_{{\bf q}\lambda}}  \int d{\bf r}\, u^{\ast} _{\sigma\tau}({\bf -r})u_{\sigma'\tau'} ({\bf - r}) {\bf p}_\lambda({\hat{\bf q}},\beta)\cdot{\partial U_{\beta}({\bf -r}-{\bf R}({\bf 0},\beta) )\over \partial {\bf R}({\bf 0},\beta)}.
\end{align}
Now, we exploit the fact that $|u_{\sigma\tau}\rangle$ is an eigenstate of the parity operator: 
$\tau^z |u_{\sigma\tau}\rangle=(-1)^\tau |u_{\sigma\tau}\rangle$.
Choosing ${\bf r}=0$ at the center of inversion, we have $u_{\sigma\tau}(-{\bf r})=(-1)^\tau u_{\sigma\tau}({\bf r})$.
Then,
\begin{align}
g^\lambda_{\sigma\tau;\sigma'\tau'}(\hat{\bf q})
=&(-1)^{\tau+\tau'}\sum_{\beta}\frac{1}{\sqrt{{\cal V}_{\rm cell}}}\sqrt{\hbar\over 2 M_{\beta}  \omega_{{\bf q}\lambda}}  \int d{\bf r}\, u^{\ast} _{\sigma\tau}({\bf r})u_{\sigma'\tau'} ({\bf  r}) {\bf p}_\lambda(\hat{\bf q},\beta)\cdot{\partial U_{\beta}({\bf -r}-{\bf R}({\bf 0},\beta) )\over \partial {\bf R}({\bf 0},\beta)}\nonumber\\
=&(-1)^{\tau+\tau'}\sum_{\beta}\frac{1}{\sqrt{{\cal V}_{\rm cell}}}\sqrt{\hbar\over 2 M_{-\beta}  \omega_{{\bf q}\lambda}}  \int d{\bf r}\, u^{\ast} _{\sigma\tau}({\bf r})u_{\sigma'\tau'} ({\bf  r}) {\bf p}_\lambda(\hat{\bf q},-\beta)\cdot{\partial U_{-\beta}({\bf -r}-{\bf R}({\bf 0},-\beta) )\over \partial {\bf R}({\bf 0},-\beta)},
\end{align}
where, in the second line, we have simply changed the dummy index $\beta$ to $-\beta$.
Now, we recognize that (i) $\beta$ and $-\beta$ are the same atom (related by the inversion operator), (ii) ${\bf R}({\bf 0},\beta)=-{\bf R}({\bf 0},-\beta)$ (because we take the origin at the center of inversion), (iii) $U_\beta(-{\bf r}+{\bf R}({\bf 0},\beta))=U_\beta({\bf r}-{\bf R}({\bf 0},\beta))$ (because $U_\beta({\bf r}-{\bf R}({\bf l},\beta))=U_\beta(|{\bf r}-{\bf R}({\bf l},\beta)|)$, (iv) ${\bf p}_\lambda({\bf q},-\beta)=(-1)^{\lambda+1} {\bf p}_\lambda ({\bf q},\beta)$, where we set the convention that $\lambda=0\, {\rm mod}\,2$ for parity-even phonon modes and $\lambda=1 \,{\rm mod}\,2$ for parity-odd phonon modes. 
This last relation emerges from the fact that, in a $q=0$ parity-even (parity-odd) mode, two atoms related by inversion have opposite (equal) displacement vectors ( see Ref.~[\onlinecite{cardona1}] ).
Then,
\begin{align}
\label{eq:pars}
g^\lambda_{\sigma\tau;\sigma'\tau'}(\hat{\bf q})
=&(-1)^{\tau+\tau'}\sum_{\beta}\frac{1}{\sqrt{{\cal V}_{\rm cell}}}\sqrt{\hbar\over 2 M_{-\beta}  \omega_{{\bf q}\lambda}}  \int d{\bf r}\, u^{\ast} _{\sigma\tau}({\bf r})u_{\sigma'\tau'} ({\bf  r}) {\bf p}_\lambda(\hat{\bf q},-\beta)\cdot{\partial U_{-\beta}({\bf -r}-{\bf R}({\bf 0},-\beta) )\over \partial {\bf R}({\bf 0},-\beta)}\nonumber\\
=&(-1)^{\tau+\tau'}\sum_{\beta}\frac{1}{\sqrt{{\cal V}_{\rm cell}}}\sqrt{\hbar\over 2  M_{\beta}\omega_{{\bf q}\lambda}}  \int d{\bf r}\, u^{\ast} _{\sigma\tau}({\bf r})u_{\sigma'\tau'} ({\bf  r}) {\bf p}_\lambda(\hat{\bf q},-\beta)\cdot{\partial U_{\beta}({\bf r}-{\bf R}({\bf 0},\beta) )\over (-)\partial{\bf R}({\bf 0},\beta)}\nonumber\\
=&(-1)^{\tau+\tau'}(-1)^{\lambda+1} (-1)\sum_{\beta}\frac{1}{\sqrt{{\cal V}_{\rm cell}}}\sqrt{\hbar\over 2 M_{\beta}  \omega_{{\bf q}\lambda}}  \int d{\bf r}\, u^{\ast} _{\sigma\tau}({\bf r})u_{\sigma'\tau'} ({\bf  r}) {\bf p}_\lambda(\hat{\bf q},\beta)\cdot{\partial U_{\beta}({\bf r}-{\bf R}({\bf 0},\beta) )\over \partial {\bf R}({\bf 0},\beta)}\nonumber\\
=&(-1)^{\tau+\tau'+\lambda} g^\lambda_{\sigma\tau;\sigma'\tau'}(\hat{\bf q}). 
\end{align}
Thus, the parity selection rule reads $\tau+\tau'+\lambda = 0\,{\rm mod}\, 2$.
In words, a parity-even $q\simeq 0$ phonon mode can only couple electronic states of the same parity, while a parity-odd $q\simeq 0$ phonon can only couple electronic states of opposite parity.
These statements are no longer exact when we keep the $i{\bf q}\cdot{\bf r}$ term in Eq.~(\ref{eq:exp2}); however, the correction due to such term is necessarily small when $q\simeq 0$.

The parity selection rule from Eq.~(\ref{eq:pars}), together with Eq.~(\ref{eq:trev}), implies that 
\begin{align}
\label{eq:gpar3}
\hat{g}^{\rm even}(\hat{\bf q}) &= g_{0 0}(\hat{\bf q}) +g_{0 z} (\hat{\bf q}) \tau^z\nonumber\\
\hat{g}^{\rm odd}(\hat{\bf q}) &= g_{0 x}(\hat{\bf q}) \tau^x +\sum_{i=x,y,z} g_{i y} (\hat{\bf q}) \sigma^i \tau^y.
\end{align}
Namely, the electronic operators appearing with parity-even phonon modes commute with $\tau^z$, while those appearing with parity-odd phonon modes anticommute with $\tau^z$.
This is the most important result from the Supplementary Material: it constitutes Eq.~(4) of the main text (N.B. in the main text we have lightened the notation slightly. For example, we have used $g_0$ rather than $g_{0 0}$, $g_z$ rather than $g_{0 z}$,  etc.). 
The coefficients appearing in Eqs.~(\ref{eq:gpar3}) can be computed from first principles with the aid of Eq.~(\ref{eq:gst}). 
For instance, $g_{0 0}=(1/4) \sum_{\sigma\tau} g^{\lambda\in\text{even}}_{\sigma\tau; \sigma\tau}$, $g_{0 z}=(1/4) \sum_{\sigma\tau} (-1)^\tau g^{\lambda\in\text{even}}_{\sigma\tau; \sigma\tau}$, $g_{0 x}=(1/4) \sum_{\sigma\tau} g^{\lambda\in\text{odd}}_{\sigma\tau; \sigma, -\tau}$, etc.

Up until now we have focused on a spatially local electron-phonon coupling.
Let us briefly comment on the nonlocal electron-phonon coupling.
To leading order, we may model the non-local electron-phonon coupling by 
\begin{equation}
\label{eq:nonloc}
{\cal H}^{\rm non-loc}_{\rm ep}=\sum_{\boldsymbol \delta}\int d{\bf r}\, \delta U({\bf r}) \psi^{\dagger}({\bf r}) \psi({\bf r}+{\boldsymbol \delta}),
\end{equation}
where ${\boldsymbol \delta}$ is the vector connecting nearest neighbor atoms.
Following the same steps outlined for ${\cal H}^{\rm loc}_{\rm ep}$, we obtain
\begin{equation}
\label{eq:nonloc2}
{\cal H}^{\rm non-loc}_{\rm ep}=\frac{1}{\sqrt{\cal V}}\sum_{{\bf k}{\bf q}\lambda}\sum_{n n'}\sum_{\boldsymbol \delta} g^\lambda_{n n'}({{\bf k,q}})e^{-i({\bf k}+{\bf q})\cdot{\boldsymbol \delta}} c_{{\bf k}n}^{\dagger} c_{{\bf k+q} n'} (a_{-{\bf q}\lambda}+a_{{\bf q} \lambda}^{\dagger}).
\end{equation}
Eqs.~(\ref{eq:loc}) and (\ref{eq:nonloc2}) are formally similar,  except for the factor $e^{-i ({\bf k}+{\bf q})\cdot{\boldsymbol\delta}}$ in the latter.
An immediate consequence of this extra factor is that $\hat{g}^\lambda$ becomes a function of both ${\bf q}$ and ${\bf k}$ rather than just ${\bf q}$, and that there are more allowed terms than those written in Eq.~(\ref{eq:gpar3}). 
Nevertheless, we are interested in weakly doped Dirac insulators, which means that the momentum of the electrons contributing to the phonon linewidths is small compared to the size of the Brillouin zone. 
This, together with the fact that $\delta$ is comparable to the inverse of the size of the Brillouin zone, implies that ${\bf k}\cdot{\boldsymbol\delta} \ll 1$.
Thus, the nonlocal electron-phonon interaction approximately reduces to the local one and Eq.~(\ref{eq:gpar3}) remains a good approximation for long-wavelength phonons.

\subsection{Real part of the phonon self-energy}

\begin{figure}
\includegraphics[width=\linewidth]{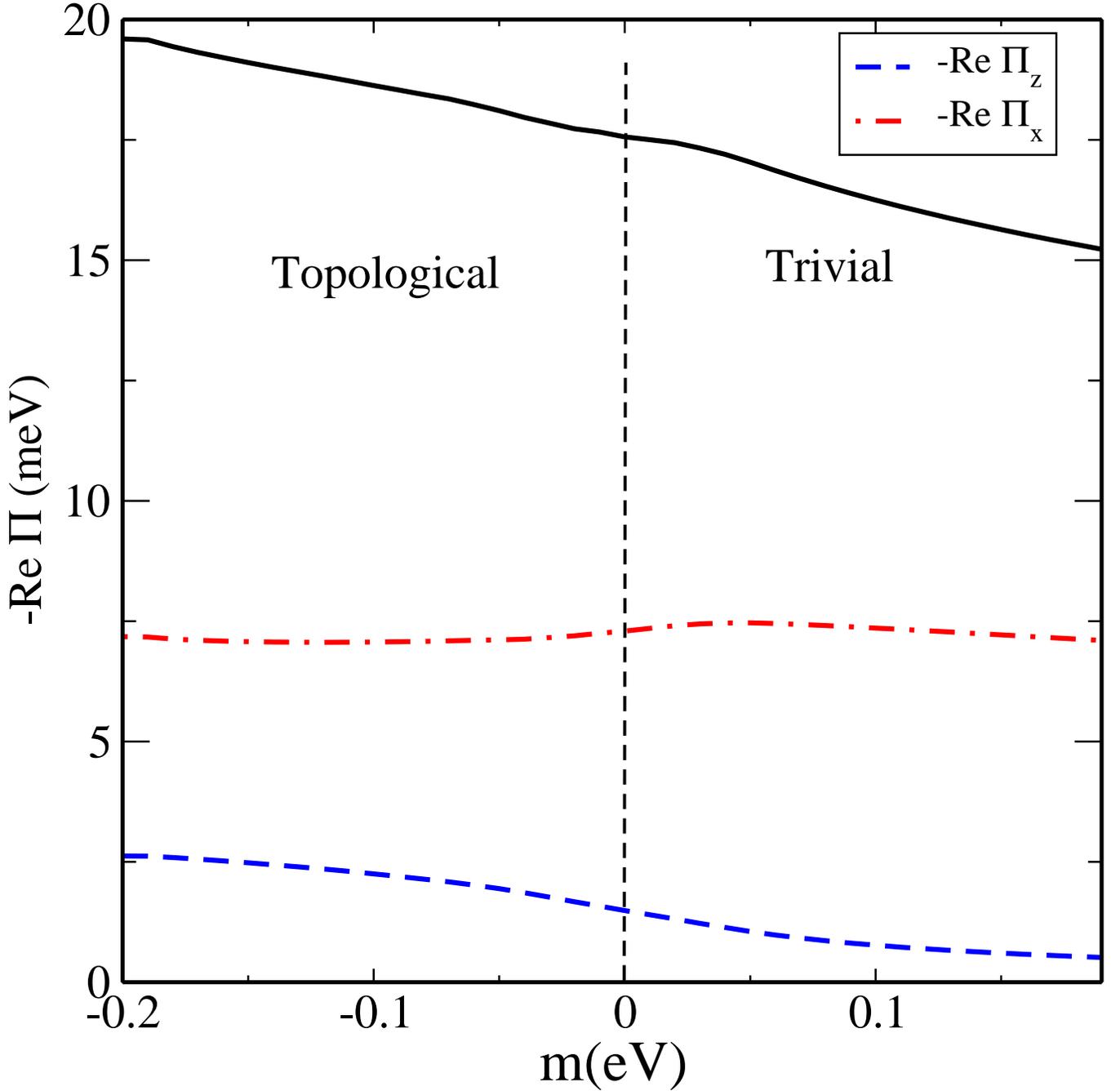}
\caption{(Color online) Real part of the electron self-energy for $q= 0$ optical phonons. 
The red (dot-dashed) and the blue (dashed) curves correspond to the ``$x$'' and ``$z$'' modes, respectively. 
The black (solid) line corresponds to the case where $g_{nn'}({\bf k},{\bf q})=1$. 
The band parameters (Sb$_2$Te$_3$) are borrowed from the Ref.~[4] of the main text. Also, $\omega_{{\bf 0}\lambda}=10\,{\rm meV}$ and $\epsilon_F=50\, {\rm meV}$.} 
\label{apnfig2}
\end{figure}

In the weak coupling regime, the electronic-induced changes in phonon frequencies are proportional to the real part of the phonon self-energy,
\begin{equation}
{\rm Re}\, \Pi_{\lambda}({\bf q},\omega_{{\bf q}\lambda})=\frac{1}{\cal V}\sum_{{\bf k},n,n'} \frac{|g^{\lambda}_{n n'}({\bf k,q})|^2(f_{{\bf k}n}-f_{{\bf k-q}n'})}
{E_{{\bf k}n}-E_{{\bf k-q}n'}-\omega_{{\bf q}\lambda}}.
\label{eq:pi}
\end{equation}
Equation~(\ref{eq:pi}) involves a sum over electron-phonon matrix elements (which contain information about the orbital character, and hence topology, of the Bloch bands) weighted with a function that depends on the details of the energy dispersion (which is only marginally related to the topological invariant).
It follows that the topological features of ${\rm Re}\,\Pi_\lambda$ are rather diluted. 
This is in stark contrast with ${\rm Im}\Pi_\lambda$, studied in the main text, where the presence of the Dirac delta gives way to energy-resolved electron-phonon matrix elements.

In support of the preceding observations, 
Fig.~(\ref{apnfig2}) illustrates ${\rm Re}\Pi_\lambda({\bf 0},\omega_{{\bf 0}\lambda})$ as a function of bare Dirac mass $m$ for the ``$x$ and $z$ modes'' discussed in the main text (which couple to electrons via $\tau^x$ and $\tau^z$, respectively). 
In addition, for reference purposes, we plot ${\rm Re}\Pi_\lambda({\bf 0},\omega_{{\bf q}\lambda})$ for a constant matrix element ($g_{n n'}^\lambda({\bf k},{\bf q}) =1$). 
The similar qualitative behavior for the ``$z$ mode'' and for the case of constant matrix elements corroborates our view that ${\rm Re} \Pi_\lambda$ is weakly sensitive to the orbital texture of the band structure. 
We also notice that the behavior of the ``$x$ mode'' is somewhat different.
This difference does originate from the electron-phonon matrix elements and does bear somewhat on the orbital character of the eigenstates; unfortunately, there are no prominent model-independent features that would allow us to learn about the band topology of a Dirac insulator by comparing the frequencies of the $x$ modes and the $z$ modes.

We conclude by stressing that ${\rm Re}\Pi_\lambda$ is smooth across a band inversion in doped samples.
This statement is in apparent disagreement with Ref.~[\onlinecite{bera}], where it has been claimed that ${\rm Re}\Pi_{\lambda}$ displays sharp features or anomalies at (or close to) a band inversion.
Moreover, we find that the non-monotonicities in ${\rm Re}\Pi_\lambda$ (e.g. its changes in slope as a function of $m$) are model-dependent and not particularly reflective of the underlying band topology.

\subsection{Imaginary part of the phonon self-energy}

In this section, we derive Eqs.~(5), (6) and (8) of the main text.
The starting point is the imaginary part of the phonon self-energy, given by
\begin{equation}
{\rm Im}\, \Pi_{\lambda}({\bf q},\omega_{{\bf q}\lambda})=\frac{\pi}{\cal V}\sum_{{\bf k} n n'} |g^{\lambda}_{n n'}({\bf k,q})|^2(f_{{\bf k}n}-f_{{\bf k-q}n'})
\delta(E_{{\bf k}n}-E_{{\bf k-q}n'}-\omega_{{\bf q}\lambda}).
\label{eq:impi}
\end{equation}
We limit ourselves to (i) long wavelength optical phonons, (ii) a four-band electronic structure with pairwise degenerate conduction and valence bands, and (iii) low temperatures.

Let us first suppose that the phonon frequency is smaller than the bulk energy gap.
In this case, $\delta(E_{{\bf k}n}-E_{{\bf k-q}n'}-\omega_{{\bf q}\lambda})$ vanishes unless $n$ and $n'$ belong to energy-degenerate bands. 
Consequently,
\begin{align}
{\rm Im}\, \Pi_{\lambda}({\bf q}\simeq {\bf 0},\omega_{{\bf q}\lambda})&\simeq\frac{\pi}{\cal V}\sum_{\substack{{{\bf k}n n'}\\ E_{{\bf k}n}=E_{{\bf k} n'}}} |g^{\lambda}_{n n'}({\bf k,\hat q})|^2 \left(f_{{\bf k}n}-f_{{\bf k}n}+\omega_{0\lambda}\frac{\partial f_{{\bf k}n}}{\partial E_{{\bf k}n}}\right)\delta({\bf v}_{{\bf k} n}\cdot {\bf q}-\omega_{{\bf 0}\lambda})\nonumber\\
&\simeq -\frac{\pi\omega_{0\lambda}}{{\cal V }}\sum_{\substack{{{\bf k}n n'}\\ E_{{\bf k}n}=E_{{\bf k} n'}}} |g^\lambda_{n n'}({\bf k,\hat q})|^2 \delta({\bf v}_{{\bf k}n}\cdot {\bf q}-\omega_{{\bf 0}\lambda})\delta(E_{{\bf k}n}-\epsilon_F),
\label{eq:impi1}
\end{align}
where ${\bf v}_{{\bf k} n}=\partial E_{{\bf k} n}/\partial {\bf k}$.
The expansion of the Fermi function in the first line of Eq.~(\ref{eq:impi1}) is valid as long as the distance between the Fermi level and bulk band edge is large compared to the phonon frequency $\omega_{{\bf 0}\lambda}$. 
Note that $q>0$ is needed in order to have a non-vanishing rhs of Eq.~(\ref{eq:impi1}).

The two bands that intersect the Fermi level are degenerate (due to inversion and time-reversal symmetry). 
Let us define their energy dispersion to be $E_{\bf k}$.
It follows that ${\bf v}_{{\bf k} n}=\partial E_{\bf k}/\partial {\bf k} \equiv {\bf v}_{\bf k}$ for both values of $n$.
Then, we may write
\begin{align}
{\rm Im}\, \Pi_{\lambda}({\bf q\simeq0},\omega_{{\bf q}\lambda}) &\simeq -\frac{\pi\omega_{0\lambda}}{{\cal V }}\sum_{\bf k} |g^\lambda_{\rm ia}({\bf k}_F,\hat{\bf  q})|^2 \delta({\bf v}_F\cdot {\bf q}-\omega_{{\bf 0}\lambda})\delta(E_{\bf k}-\epsilon_F)\nonumber\\
&\equiv-\pi\omega_{{\bf 0}\lambda} D(\epsilon_F) \overline{|g^\lambda_{\rm ia}({\bf k}_F, \hat{\bf q})|^2 \delta({\bf v}_F\cdot{\bf q}-\omega_{{\bf 0}\lambda})},  
\label{eq:impi2}
\end{align}
where ${\bf v}_F\equiv {\bf v}_{{\bf k}_F}$ is the Fermi velocity, $D(\epsilon_F)=\frac{1}{\cal V}\sum_{\bf k}\delta(E_{\bf k}-\epsilon_F)$ is the density of states per band at the Fermi level, 
\begin{equation}
|g^\lambda_{\rm ia} ({\bf k}_F,{\bf q})|^2\equiv\sum_{n n'} |g^{\lambda}_{n n'}({\bf k,q})|^2\,\,\,\text{    (with $E_{{\bf k}n}=E_{{\bf k} n'}=\epsilon_F$)},
\end{equation}
and
\begin{align}
\overline O\equiv \sum_{\bf k} O_{\bf k}\delta(E_{\bf k}-\epsilon_F)/\sum_{\bf k} \delta(E_{\bf k}-\epsilon_F)
\end{align}
is the Fermi-surface average of $O$.
Hence, Eq.~(5) in the main text is obtained via the relation $\gamma_{\rm ia}^{\lambda}=-{\rm Im}\, \Pi_{\lambda}({\bf q\simeq0},\omega_{{\bf q}\lambda})$. 

From Eq.~(\ref{eq:impi2}), it is straightforward to derive Eq.~(6) of the main text.
For this derivation, we adopt the model Hamiltonian of Eq.~(1) in the main text, which is spherically symmetric.
Accordingly, we may write
\begin{align}
{\rm Im}\, \Pi_{\lambda}({\bf q\simeq 0},\omega_{{\bf q}\lambda})&\simeq -\frac{\pi\omega_{0\lambda}}{{\cal V } q v_F}\sum_{{\bf k}} |g^{\lambda}_{\rm ia}({\bf k,\hat q})|^2 \delta(\hat{\bf v}_F\cdot {\bf \hat q}-\frac{\omega_{{\bf 0}\lambda}}{q v_F})\delta(E_{{\bf k}}-\epsilon_F)\nonumber\\
&=-\frac{\pi\omega_{0\lambda}}{q v_F}\int_0^\infty \frac{dk k^2}{(2\pi)^3}\int_{-1}^{1} d(\cos\theta)\int_0^{2\pi} d\varphi\, |g^{\lambda}_{\rm ia}({\bf k,\hat q})|^2 \delta\left(\cos\theta-\eta\right)\delta(E_{\bf k}-\epsilon_F),
\end{align}
where $\hat{\bf v}_F={\bf v}_F/v_F $, $\hat{\bf q}={\bf q}/q $ and $\eta=\omega_{{\bf 0}\lambda}/(q v_F)$.
At this point, we are interested in phonon modes that couple to electrons through $\tau^x$ or $\tau^z$ (cf. the paragraph preceding Eq.~(6) in the main text).
For these, we can write
\begin{equation}
|g^\lambda_{\rm ia} ({\bf k},{\bf q}\simeq {\bf 0})|^2 = |g_j (\hat{\bf q})|^2 \sum_{\substack{{n n'}\\ E_{{\bf k}n}=E_{{\bf k} n'}}}  |\langle u_{{\bf k} n}|\tau^j|u_{{\bf k} n'}\rangle|^2 \equiv |g_j (\hat{\bf q})|^2 |\langle\tau^j\rangle_{\rm ia}|^2,
\end{equation}
where the sum over $n$ and $n'$ is constrained to the two degenerate bands that intersect the Fermi surface, and  $j$ equals either $x$ or $z$ (N.B. there is no sum over $j$).
In the spherically symmetric model,  $|\langle\tau^j\rangle_{\rm ia}|^2$ depends only on the magnitude of ${\bf k}$.
Consequently,
\begin{align}
{\rm Im}\, \Pi_j({\bf q\simeq 0},\omega_{{\bf q}j})
&=-\frac{\pi\omega_{{\bf 0}j} |g_j(\hat{\bf q})|^2}{q v_F} 2\pi \int_0^\infty \frac{ k^2 dk}{(2\pi)^3}  |\langle \tau^j\rangle_{\rm ia}|^2  \delta(E_{{\bf k}}-\epsilon_F)\int_{-1}^1 d(\cos\theta)\, \delta\left(\cos\theta-\eta\right)\nonumber\\
&=-\frac{\pi\omega_{{\bf 0}j} |g_j(\hat{\bf q})|^2}{q v_F^2} 2\pi \int_0^\infty \frac{ k^2 dk}{(2\pi)^3}  |\langle\tau^j\rangle_{\rm ia}|^2 \delta(k-k_F)\int_{-1}^1 d(\cos\theta)\,\delta\left(\cos\theta-\eta\right)\nonumber\\
&=-\frac{\pi\omega_{{\bf 0}j} |g_j(\hat{\bf q})|^2}{q v_F^2} 2\pi\overline{|\langle\tau^j\rangle_{\rm ia}|^2} \frac{k_F^2}{(2\pi)^3}  \int_{-1}^1 d(\cos\theta)\,\delta\left(\cos\theta-\eta\right)\nonumber\\
&=-\pi\eta |g_j(\hat{\bf q})|^2 \overline{|\langle\tau^j\rangle_{\rm ia}|^2}  \frac{ 2\pi k_F^2 }{ (2\pi)^3 v_F} \Theta(1-\eta)\nonumber\\
&=-\pi\eta |g_j(\hat{\bf q})|^2  \overline{|\langle\tau^j\rangle_{\rm ia}|^2} \frac{D(\epsilon_F)}{2}\Theta(1-\eta),
\label{eq:e6}
\end{align}
where $\Theta(x)$ is the step function and
\begin{equation}
D(\epsilon_F)=\frac{1}{\cal V}\sum_{{\bf k} }\delta(E_{{\bf k}}-\epsilon_F)=\frac{4\pi}{(2\pi)^3}\frac{k_F^2}{v_F}.
\end{equation}
By taking the negative of Eq.~(41), we obtain Eq.~(6) of the main text.

Finally, we derive Eq.~(8) of the main text.
In order to do so, we consider the case where the phonon frequency exceeds the bulk energy gap. 
Then, 
\begin{equation}
{\rm Im}\, \Pi_{\lambda}({\bf q\simeq 0},\omega_{{\bf q}\lambda})\simeq\frac{\pi}{\cal V}\sum_{{\bf k}, n\in c, n'\in v} |g^{\lambda}_{n n'}({\bf k,\hat q})|^2 (f_{{\bf k}n}-f_{{\bf k}n'})\delta(E_{{\bf k}n}-E_{{\bf k}n'}-\omega_{{\bf 0}\lambda}),
\end{equation}
where $c$ and $v$ correspond to conduction and valence bands, respectively.
Defining the respective energies as $E_{{\bf k} c}$ and $E_{{\bf k} v}$, and assuming for simplicity that the Fermi energy lies inside the Dirac gap, we have
\begin{align}
{\rm Im}\, \Pi_{\lambda}({\bf q\simeq 0},\omega_{{\bf q}\lambda})&\simeq-\frac{\pi}{\cal V}\sum_{{\bf k}} |g^{\lambda}_{cv}({\bf k,\hat q})|^2\delta(E_{{\bf k}c}-E_{{\bf k}v}-\omega_{{\bf 0}\lambda})\nonumber\\
&\equiv-\pi D_{\rm joint}(\omega_{{\bf 0}\lambda})\overline{|g^{\lambda}_{\rm ie}({\bf k,\hat q})|^2},
\end{align}
where $|g^\lambda_{\rm ie}|^2\equiv\sum_{n\in{\rm c}, n'\in{\rm v}} |g^\lambda_{n n'}|^2$,
\begin{align}
D_{\rm joint}(\omega_{0\lambda})=\frac{1}{\cal V} \sum_{\bf k}\delta(E_{{\bf k}c}-E_{{\bf k}v}-\omega_{{\bf 0}\lambda})
\end{align}
is the joint density of states and
\begin{align}
\overline{|g^{\lambda}_{\rm ie}({\bf k,\hat q})|^2}=\sum_{\bf k} |g^{\lambda}_{\rm ie}({\bf k,\hat q})|^2\delta(E_{{\bf k}c}-E_{{\bf k}v}-\omega_{{\bf 0}\lambda})/\sum_{\bf k}\delta(E_{{\bf k}c}-E_{{\bf k}v}-\omega_{{\bf 0}\lambda}).
\end{align}  
Thus, we recover Eq.~(8) in the main text.

 \end{widetext}

\end{document}